\documentclass[iop]{emulateapj}        
\usepackage{epsfig}
\usepackage{epstopdf}            
\usepackage{graphicx,color}        
\usepackage{amssymb}            
\usepackage{color}            
\usepackage{url}            
\usepackage{amsmath}            
\usepackage{rotating}            
\usepackage{float}            
\usepackage{textcomp}            
\usepackage{dcolumn}
\usepackage{times}
\usepackage{tabularx}
\usepackage[colorlinks=true,citecolor=blue]{hyperref}
\usepackage[english]{babel}
\usepackage{amssymb}
\usepackage{pifont}
\newcommand{\cmark}{\ding{51}}%
\newcommand{\xmark}{\ding{55}}%

\shorttitle{Supersonic downflows above sunspots}
\shortauthors{T. Samanta et al.}

\begin{document}
\title{Statistical investigation of supersonic downflows in the transition region above sunspots} 

\author{
Tanmoy Samanta$^{1}$,
Hui Tian$^{1}$,
Debi Prasad Choudhary$^{2}$  
           }
\affil{             $^{1}$ School of Earth and Space Sciences, Peking University, Beijing 100871, China;
                      {\color{blue}{tsamanta@pku.edu.cn; huitian@pku.edu.cn}}\\
                      $^{2}$ Department of Physics \& Astronomy, California State University, Northridge, CA 91330-8268, USA}

\begin{abstract}
Downflows at supersonic speeds have been observed in the transition region (TR) above sunspots for more than three decades. These downflows are often seen in different TR spectral lines above sunspots. We have performed a statistical investigation of these downflows using a large sample which was missing earlier. The Interface Region Imaging Spectrograph (IRIS) has provided a wealth of observational data of sunspots at high spatial and spectral resolution in the past few years. We have identified sixty datasets obtained with IRIS raster scans. Using an automated code, we identified the locations of strong downflows within these sunspots. We found that around eighty percent of our sample show supersonic downflows in the \ion{Si}{4} 1403~{\AA} line. These downflows mostly appear in the penumbral regions, though some of them are found in the umbrae. We also found that almost half of these downflows show signatures in chromospheric lines. Furthermore, a detailed spectral analysis was performed by selecting a small spectral window containing the \ion{O}{4} 1400/1401~{\AA} and \ion{Si}{4}1403~{\AA} lines. Six Gaussian functions were simultaneously fitted to these three spectral lines and their satellite lines associated with the supersonic downflows. We calculated the intensity, Doppler velocity and line width for these lines. Using the \ion{O}{4} 1400/1401~{\AA} line ratio, we find that the downflow components are around one order of magnitude less dense than the regular components. Results from our statistical analysis suggest that these downflows may originate from the corona and that they are independent of the background TR plasma.  
\end{abstract}

\section{Introduction}
\begin{figure*}[!htbp]
\centering
\includegraphics[angle=0,clip,width=18.1cm]{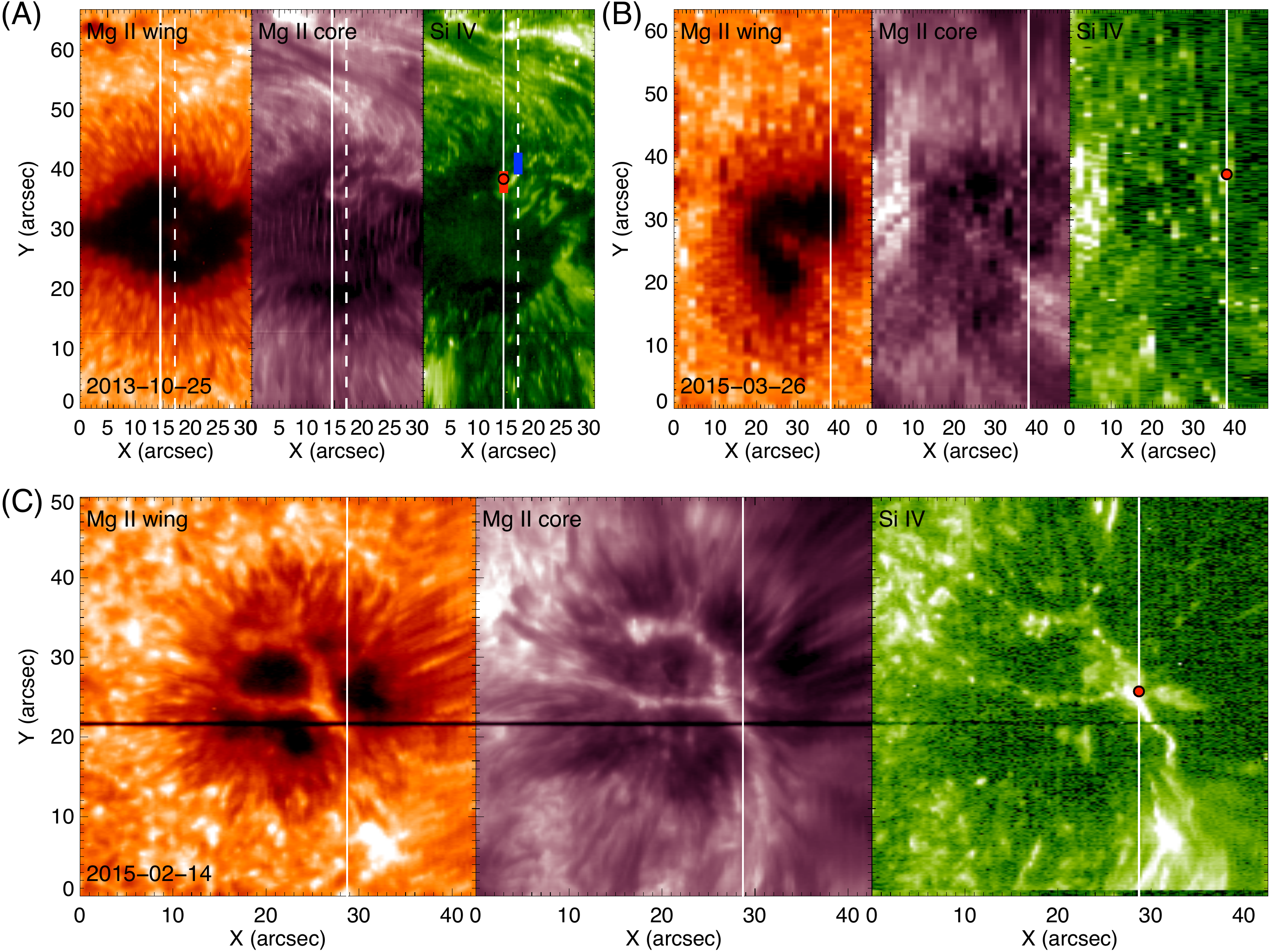}
\caption{Panel (A) shows \ion{Mg}{2}~k wing, \ion{Mg}{2}~k core and \ion{Si}{4} core intensity images of a sunspot observed on  2013 October 25.  The \ion{Mg}{2}~k wing image is constructed by integrating the emission over a 1~{\AA} wavelength window around 2795~{\AA}, whereas \ion{Mg}{2}~k core and \ion{Si}{4} core images are created by integrating the emission over a 0.4~{\AA} wavelength window around the line center.
The IRIS spectra along the solid and dashed lines are shown in Figures~\ref{fig2} and \ref{fig3}, respectively. 
The red and blue stripes above the solid line and dashed line, respectively, indicate the blue and red area marked in the Figures~\ref{fig2} and \ref{fig3}. 
(B)--(C): Two other sunspots as observed on 2015 March 26 and 2015 February 14, respectively. The panels are the same as in (A). 
A small window of the IRIS spectra containing the doublet \ion{O}{4} and \ion{Si}{4} lines along the vertical solid line in (A), (B) and (C) are shown in Figure~\ref{fig5}. The line profiles and their Gaussian fits at the locations of the red dots enclosed with black circles are also shown in Figure~\ref{fig5}. Particularly, these positions represent the locations of the dashed lines as marked in the spectra shown in the left panels of Figure~\ref{fig5}.}
\label{fig1} 
\end{figure*}
\begin{figure*}
\centering
\includegraphics[angle=0,clip,width=18.3cm]{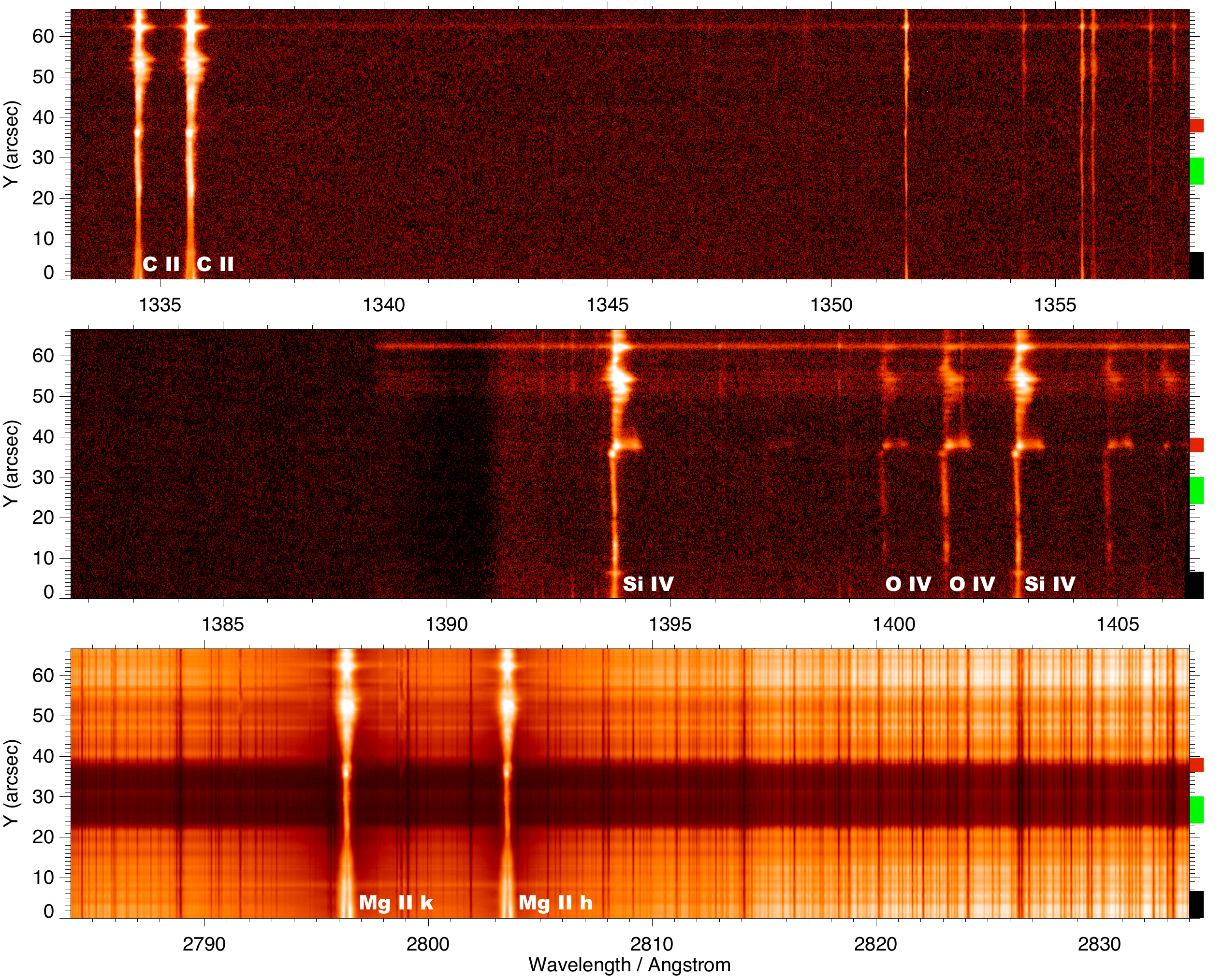}
\caption{ The IRIS spectra along the solid line shown in Figure~\ref{fig1}(A). Top to bottom panels show the IRIS spectra in the short-wavelength far ultra-violet (FUVS), long-wavelength far ultra-violet (FUVL) and near ultra-violet (NUV) bands. The spectra averaged over the red, green and black regions marked at the right side of each panel are shown in Figure~\ref{fig4}, where we also label each spectral line for reference.}
\label{fig2} 
\end{figure*}

Sunspots, regions of concentrated strong magnetic fields in the surface of the Sun, appear as dark features in the photosphere due to suppressed convection. The central region of a simple sunspot is usually dark and known as the umbra, which is generally surrounded by a less dark region named penumbra. 
Most earlier studies of sunspots are focused on the photosphere/chromosphere using different high-resolution ground-based telescopes and space-based observatories \citep{2003A&ARv..11..153S,2011LRSP....8....4B,2017arXiv171207174T}. 
After the launch of the Solar and Heliospheric Observatory (SOHO) in 1995, observations by the Solar Ultraviolet Measurements of Emitted Radiation  \citep[SUMER;][]{1995SoPh..162..189W} and Coronal Diagnostic Spectrometer \citep[CDS;][]{Harrison95} instruments on board SOHO explored the dynamics of sunspots in the transition region (TR).  Unfortunately, most of the sunspot observation campaigns with SUMER were performed during the initial couple of years after the lunch of SOHO, when the Sun was less active, and there were fewer sunspots. The CDS instrument obtained spectra of some strong TR lines for many years. However, the spatial resolution was low compared to the SUMER instrument.

The Interface Region Imaging Spectrograph \citep[IRIS;][]{2014SoPh..289.2733D} has been providing a wealth of high-resolution data since 2013, allowing us to investigate the highly dynamic and less understood chromosphere and TR. Sunspot studies in the TR have received a significant revolution in the IRIS-era. Unprecedented observations of sunspots by IRIS have unveiled several new types of small-scale dynamic events in the TR above sunspots, e.g., penumbral bright dots \citep{2014ApJ...790L..29T,2015ApJ...811L..33V,2016ApJ...822...35A,2017ApJ...835L..19S}, UV bursts around sunspots and light bridges \citep{2015ApJ...811..137T,2018ApJ...854...92T,2018ApJ...855L..19R}. For an overview of these newly discovered phenomena in the TR above sunspots, we refer to \citet{2017RAA....17..110T} and \citet{Tian2018}.

Here we study a particular phenomenon in the TR above sunspots: supersonic downflows.
Supersonic downflows were first observed in the spectra of TR lines acquired by the High-Resolution Telescope
and Spectrograph (HRTS) \citep{1982SoPh...77...77D,1982SoPh...81..253N, 1990Ap&SS.170..135B,1993ApJ...412..865G,1988ApJ...334.1066K}.
These flow are occasionally seen above sunspots in TR spectral lines which are formed around a temperature of 10$^{5.0}$ K. These spectral lines often show two components, one is nearly stationary, and the other is redshifted with a $\sim$100 km~s$^{-1}$ velocity. As the flow speed is higher than the sound speed at such a temperature, these flows are called supersonic downflows. 
Later, such supersonic downflows were also observed by SUMER \citep{2001ApJ...552L..77B, 2004ApJ...612.1193B}.  \citet{2001ApJ...552L..77B} analyzed SUMER sit-and-stare observations and found that profiles of emission lines from the ions of \ion{N}{5} and \ion{O}{5} can be represented by two Gaussian functions at the locations of supersonic downflows. 
They named these flows ``dual flows"  because the primary component conjointly shows a small red shift additionally to the second component with a supersonic speed.
\citet{2004ApJ...612.1193B} extended their earlier work and performed a study of SUMER raster scans of
12 sunspots and found that dual flows exist in five out of them. And most of these flows appear to exist in the penumbrae.

Recently, after the launch of the IRIS mission, supersonic downflows in TR lines above a
sunspot were detected by \citet{2014ApJ...789L..42K}. The observed downflows appear to be bursty in nature and show extremely broad line profiles in the \ion{Mg}{2}~h and k, \ion{C}{2} 1336~{\AA}, \ion{Si}{4} 1394~{\AA} and 1403~{\AA}~lines, 
with downflow velocities of up to 200 km~s$^{-1}$.  It appears that the properties of these downflows are very different from what was reported earlier by HRTS and SUMER. 
\citet{2016ApJ...823...60B, 2016ApJ...829..103D, 2017ApJ...835L..19S} also found signatures of downflows at the locations of short-lived subarcsecond bright dots \citep{2014ApJ...790L..29T} in the TR. It appears that these transient downflows are not supersonic, thus are different from the steady supersonic downflows observed earlier with HRTS and SUMER.
These transient downflows could be due to falling plasma into sunspots which are sometimes named coronal rain. \citet{2015A&A...582A.116S} investigated a supersonic downflow event above a sunspot using a sit-and-stare observation of IRIS. 
They found that the downflow component observed in several \ion{Si}{4} and \ion{O}{4} lines is redshifted by around 90 km~s$^{-1}$ and stays steady for almost 80 min during the total period of the IRIS sit-and-stare observation.  They also reported that the downflow occurs in the umbra, and shows no signature in the chromosphere. The detailed analysis of the event is summarized in \citet{2015A&A...582A.116S}, where they determined the size, density, mass flux of this single downflow event. 
Such a supersonic downflow appears to show no signature of oscillation, as mentioned by \citet{2014ApJ...786..137T}. \citet{2016A&A...587A..20C} analyzed an IRIS raster observation and tried to relate the TR dynamics with coronal structures using the data from the Atmospheric Imaging Assembly \citep[AIA;][]{2012SoPh..275...17L} on board the Solar Dynamics Observatory. They found that high-speed downflows from the upper atmosphere could heat the dense plasma at the footpoints of coronal loops. 
\citet{2015A&A...582A.116S} and \citet{2016A&A...587A..20C} suggested that these persistent supersonic downflows may produce stationary shocks which could heat the plasma. Previous observations show that these supersonic downflows are usually seen in emission lines formed only in the middle of the TR, where the temperature is in the range of 10$^{4.5}$~K--10$^{5.5}$~K. Reported observations generally do not show signatures of these supersonic downflows in the chromosphere.  
Regarding these supersonic downflows, several questions remain to be answered. For instance, how common are these downflows? Do they also show signatures in chromosphere lines? Are these downflows mostly common in penumbral regions? 
What are the densities and mass fluxes of these downflows? 
Are these downflows locally generated in the TR or coming from a higher height? 
To help answer these questions, a statistical investigation of these downflows is highly desired.

In this work, we present results from a detailed statistical study of the properties of these supersonic downflows using the IRIS data. 
In Sect. 2, we describe the details of observations and data selection criteria. The analysis methods and
results are presented in Sect. 3. A discussion and the summary are made in Sect. 4 and 5, respectively.

%

\begin{figure*}
\centering
\includegraphics[angle=0,clip,width=18.3cm]{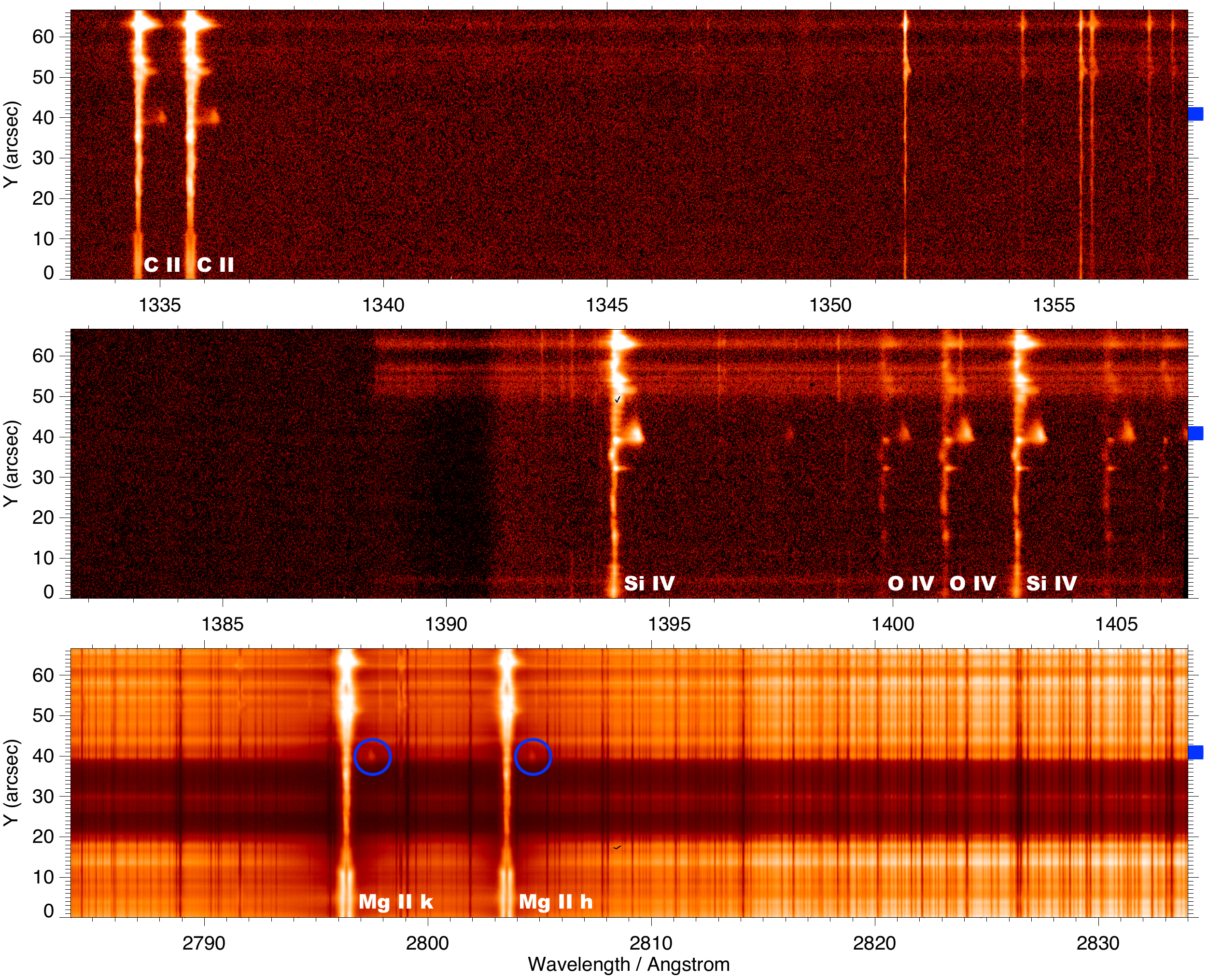}
\caption{The IRIS spectra along the dashed line shown in Figure~\ref{fig1}(A). The panels are similar to those in Figure~\ref{fig2}. The spectrum averaged over the blue region marked at the right side of each panel is shown in Figure~\ref{fig4}. The two small circles in the NUV spectral window mark the downflow components in the \ion{Mg}{2}~h/k lines.}
\label{fig3} 
\end{figure*}
\begin{figure*}
\centering
\includegraphics[angle=0,trim = 0mm 0mm 0mm -5mm,clip,width=18.3cm]{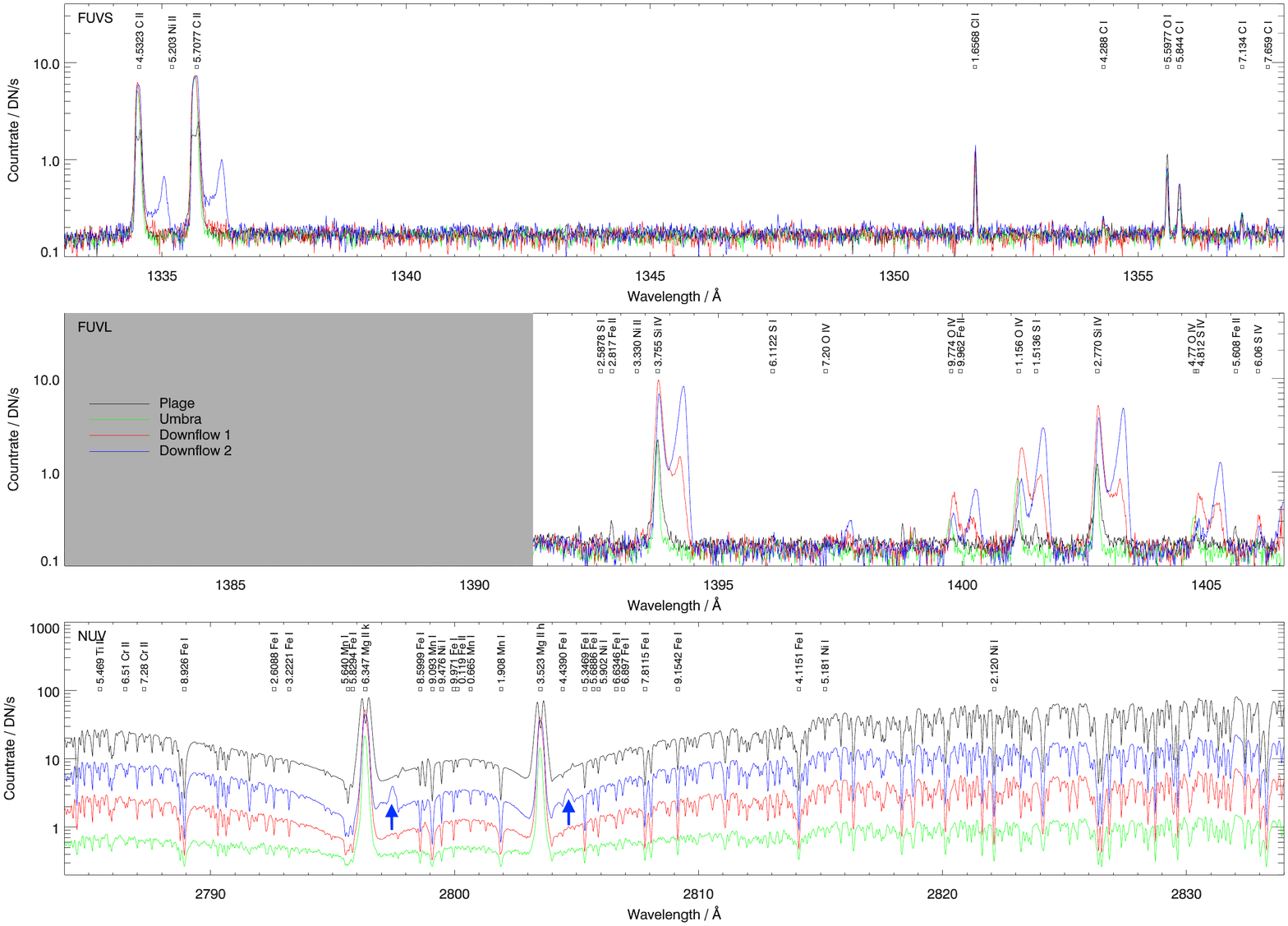}
\caption{Average IRIS spectra at specific locations of two slits as marked in Figures~\ref{fig2} and \ref{fig3}. 
The black lines represent the spectra of typical plage region, whereas the green lines represent the spectra of a typical umbral region (marked on the right side of each panel of Figure~\ref{fig2}). The red and blue lines are the average  spectra of two regions where supersonic downflows are observed. These two regions are marked at the right side of each panel of Figure~\ref{fig2} and \ref{fig3}, respectively.  The red spectra show a downflow only in TR lines, whereas the blue spectra show a downflow in TR lines as well as in the chromospheric  \ion{C}{2} and \ion{Mg}{2}~h/k lines.
Two arrows in the NUV spectral window mark the downflow components seen in the \ion{Mg}{2}~h/k lines (spectra shown in blue). For the identification of different lines, we refer to \citet{2017RAA....17..110T}.}
\label{fig4} 
\end{figure*}

\section{Observations}

IRIS has observed hundreds of sunspots over the past few years, including raster and sit-and-stare observations. In this work, we use the raster observations of sunspots to perform a statistical investigation of supersonic downflows. We have chosen 60 raster observations of sunspots from 2013 September 01 to 2015 April 01 for our analysis. The primary selection criteria are the following:  a) We first chose the observations where the slit scans a full sunspot or most of a sunspot. b) We excluded pores and sunspots/active regions with flaring activities. c) We also limited our observations close to disk center (distance to disk center $\le$ 600$''$). d)  If there are many repeated raster scans of the same sunspot, we randomly selected one raster.  A table containing the list of observational data is provided in the Appendix. The table provides some information such as the active region numbers, observation times, and locations of the supersonic downflows.

We have used the Level~2 data (available at \url{http://iris.lmsal.com/search}), where the dark current, flat field and geometrical and wavelength corrections have all been applied. We have made movies of each sunspot spectra for visual inspection of supersonic downflows. Three movies corresponding to the observations presented in Figure~\ref{fig1} are available online. Out of the 60 observations, 48 reveal signatures of supersonic downflows. We find that among these 48 observations, supersonic downflows mostly occur in the penumbrae in 28 observations. In only 4 observations the downflows appear mainly in the umbrae. In the rest 16 observations, signatures of downflows are found in both the umbrae and penumbrae. We have provided this information in the table in the Appendix.
We have also carefully examined the movies of the rasters showing no sign of downflows and found that the sunspots in these observations are mostly very simple circular-type and small in size (seven of them). Two of the rasters have very low exposure times and might not have enough signal for the detection of downflows. Three of them are complex and large, but no supersonic downflows were detected.

\begin{figure*}[!htbp]
\centering
\includegraphics[angle=00,clip,width=18.4cm]{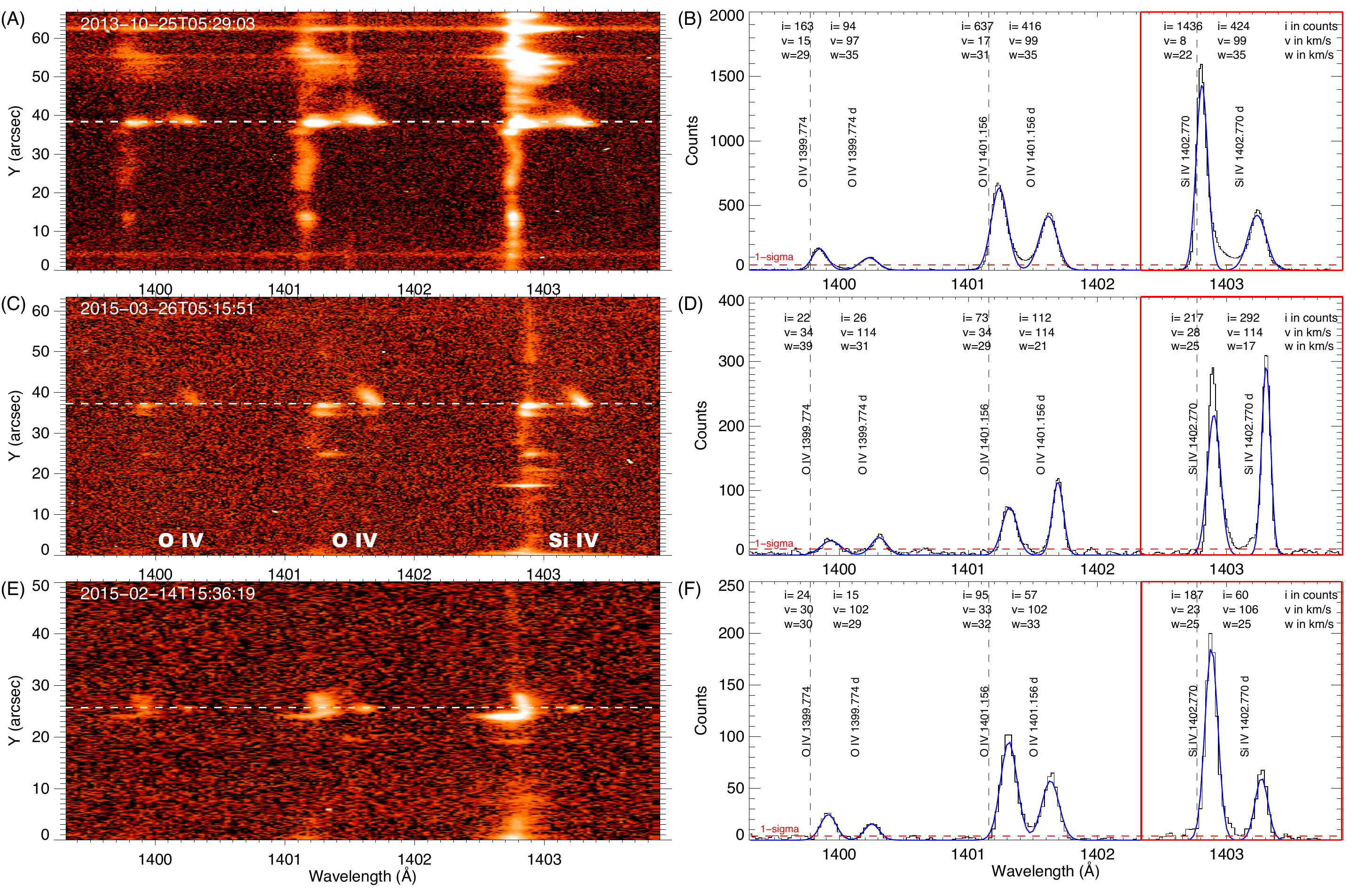}
\caption{ IRIS spectra containing two \ion{O}{4} lines and one \ion{Si}{4} line acquired along the vertical solid lines marked in Figure~\ref{fig1}. The spectral line profiles at the location indicated by the dashed line (also shown with the red dots in Figure~\ref{fig1}) of each left panel are shown in the corresponding right panel. A six-component Gaussian fit to the spectra of the regular and downflow components of the \ion{O}{4} 1400/1401~{\AA} and \ion{Si}{4} 1403~{\AA} lines is shown as the overplotted blue line in each right panel. The fitting parameters (i--peak intensity, v--Doppler shift, w--line width) are also printed. The spectra within the red square box contain the \ion{Si}{4}  line and its downflow component, which are used for initial detection of the supersonic downflows (see the text for details). }
\label{fig5} 
\end{figure*}
%
\begin{figure*}
\centering
\includegraphics[angle=00,clip,width=18.5cm]{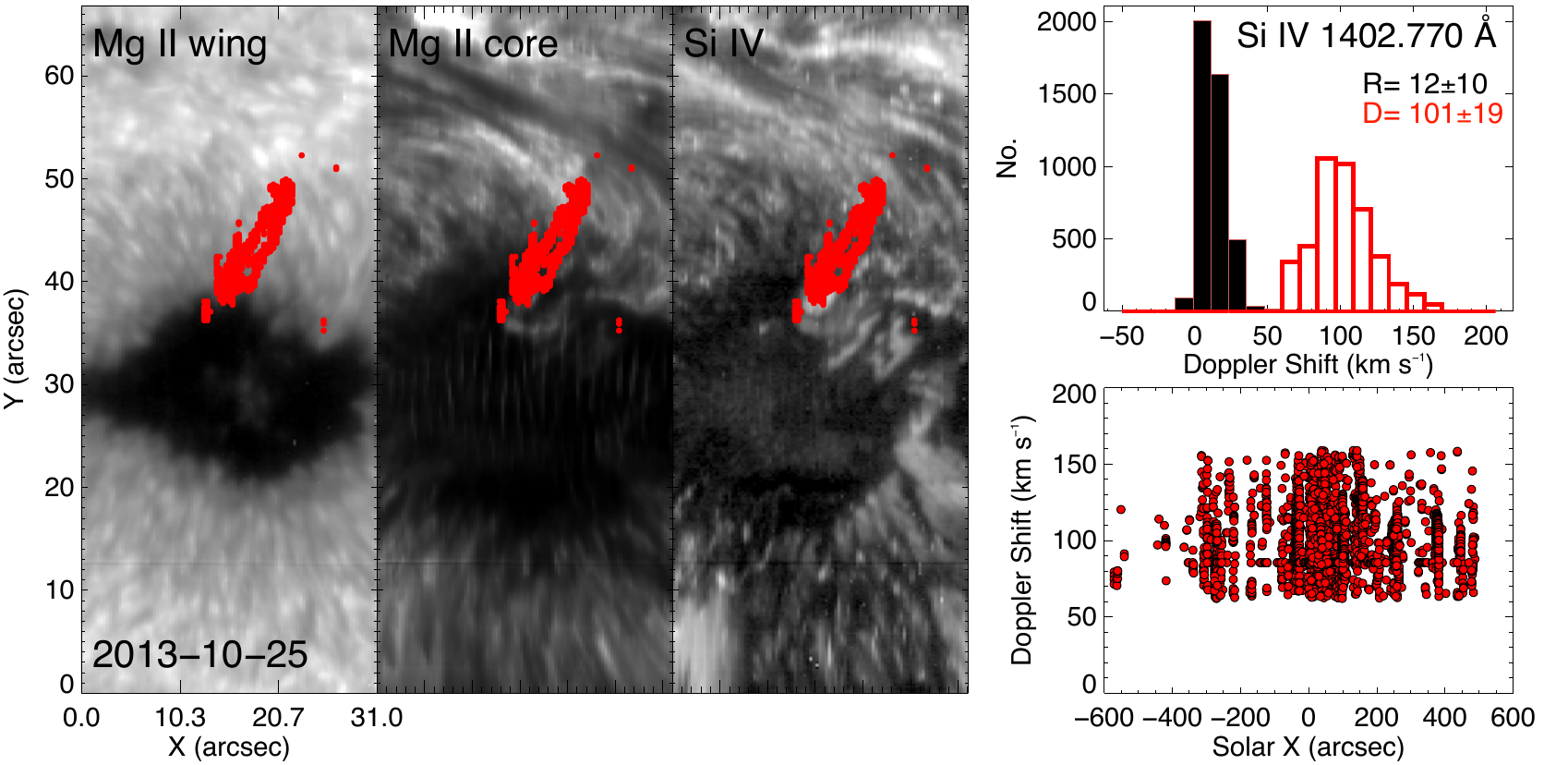}
\caption{Left three panels: locations of detected downflows are shown as overplotted red dots in the \ion{Mg}{2} and \ion{Si}{4} images of the same sunspot as in Figure~\ref{fig1}(A). 
These detected locations are based on the analysis of the \ion{Si}{4} line only.
Right top panel: The distribution of the Doppler shift as measured from only \ion{Si}{4} spectral window (a total of 4,976-pixel locations, please see Sec.~\ref{fit_spectra} for details).
The black and red histograms correspond to the regular and downflow components, respectively. The average values of the velocities with 1$\sigma$ errors are also printed in the panel. Right bottom panel: The scatter plot of the relationship between the Doppler velocity of the downflow component and the solar-x.}
\label{fig6} 
\end{figure*}
\section{Data Analysis and Results} 
\subsection{IRIS Rasters and Spectra} 

IRIS has three spectral channels: one near ultraviolet
(NUV) channel with a wavelength coverage of  2782--2835~{\AA}, and two far ultraviolet (FUV) channels covering 1332--1358~{\AA} (FUVS) and 1389--1407~{\AA} (FUVL) \citep{2014SoPh..289.2733D}. The NUV channel covers two strong spectral lines formed in the chromosphere (\ion{Mg}{2}~h 2803~{\AA} and \ion{Mg}{2}~k 2796 ~{\AA}), and the FUVL channel can record several TR lines such as the \ion{Si}{4} 1394/1403~{\AA} and \ion{O}{4} 1400/1401~{\AA} lines. The FUVS channel includes the strong \ion{C}{2} 1334/1335~{\AA} lines which are formed in the upper chromosphere or lower TR. The \ion{Mg}{2}~h 2803~{\AA} and \ion{Mg}{2}~k 2796 ~{\AA} lines are formed around a temperature of 10,000 K, whereas the emission at their wings primarily comes from regions with a temperature of 5,000--8,000 K.  The  \ion{C}{2} 1334/1335~{\AA}  lines sample the plasma with a temperature of around 25,000 K. The \ion{Si}{4} and \ion{O}{4} TR lines form at around 80,000K and 150,000 K, respectively. Figure~\ref{fig1} shows raster images of three different sunspots.  The \ion{Mg}{2}~k wing, \ion{Mg}{2}~k core and \ion{Si}{4} intensity images are shown for each observation.
The IRIS spectra along the solid line in Figure~\ref{fig1}(A) are shown in Figure~\ref{fig2}. Secondary components of the TR lines (see the middle panel which contains the \ion{Si}{4} and \ion{O}{4} lines) are clearly seen at some locations of the sunspot. The spectra at the top panel (FUVS channel containing the \ion{C}{2} 1334/1335~{\AA} lines) and also the bottom panel (NUV channel containing the \ion{Mg}{2}~h 2803~{\AA} and \ion{Mg}{2}~k 2796 ~{\AA} lines) do not show any visible signatures of the Doppler-shifted secondary component at the same locations.

The spectra along the dashed line marked in the Figure~\ref{fig1}~(A) are shown in Figure~\ref{fig3}. We can clearly observe the presence of secondary components in the TR lines. Additionally, we notice that at the location of the supersonic downflow, the chromospheric \ion{C}{2} 1334/1335~{\AA} lines, \ion{Mg}{2}~h 2803~{\AA} and \ion{Mg}{2}~k 2796 ~{\AA}  lines reveal a clear presence of a redshifted component.
Hence, it is clear that at some locations the TR supersonic downflows show no distinct signatures in chromospheric lines (Fig.~\ref{fig2}). Whereas other supersonic downflows have strong footprints at lower temperatures (Fig.~\ref{fig3}). In Figure~\ref{fig4},  we compare the average spectra of the plage region, normal sunspot region without a signature of supersonic downflow and two different supersonic downflows: one showing no signature and the other showing prominent signature in the chromospheric lines.  
Manual inspection of all the movies as described in the observation section shows that out of the 48 observations where supersonic downflows are present, 22 show some signs of downflows in the chromospheric lines. These signs include not only a brightening but also a clear Doppler-shifted component in either the \ion{C}{2} or \ion{Mg}{2} lines or both. 
In our future study, we plan to investigate the chromospheric and coronal associations of the supersonic downflows in detail using data taken by multiple instruments.
In this work, we focus our study on the \ion{Si}{4} 1403~{\AA} and \ion{O}{4} 1400/1401~{\AA} lines. 

\subsection{Detection of supersonic downflows and multi-Gaussian fitting to their Spectra}
\label{fit_spectra}
To derive spectroscopic parameters of the supersonic downflows, we have selected a small spectral window of 1399.31~{\AA}--1403.89 ~{\AA}, which contains the \ion{O}{4} 1400/1401~{\AA} doublet and \ion{Si}{4} 1403~{\AA} spectral lines. The primary reason for selecting this wavelength range is that \ion{Si}{4}1403~{\AA} is one of the strongest TR lines in the IRIS spectral range. Also, the  \ion{O}{4} 1400/1401~{\AA} doublet lines are useful to determine the electron density of the TR plasma. Figure~\ref{fig5} presents some examples of the spectra in the selected window. A total of six isolated components corresponding to the three spectral lines \ion{O}{4} 1400~{\AA}, \ion{O}{4} 1401~{\AA} and \ion{Si}{4} 1403~{\AA} and their satellite lines (due to supersonic downflows) can be seen at some locations.
To increase the signal to noise (S/N), we have made a 3-point running smoothing along the wavelength and also along the slit dimensions.

The detection of supersonic downflows is sometimes difficult as the signal within sunspots is very weak (though it also depends on the exposure, binning of the spectra, etc.). Also, the TR is very dynamic and often shows broad line profiles due to different activities. To ensure that we primarily detect supersonic downflows, we use a few selection criteria. 
For initial detection, we have first concentrated only on the \ion{Si}{4} line as it usually has a higher signal than the \ion{O}{4} lines. This narrow spectral window is shown in Figure~\ref{fig5}(B) with a red rectangular box.  First, we determined the location and amplitude of the \ion{Si}{4} 1403~{\AA} line. Afterward, we examined if there are any signs of a secondary component beyond 40 km~s$^{-1}$ (also less than 160 km~s$^{-1}$) on the red side. If there is any signal of a Doppler shifted component with an amplitude at least 20 percent of the primary component, then we count it as an initial detection of supersonic downflow.  A similar condition was also applied by \citet{2001ApJ...552L..77B}. With this initial condition, we detected 5,391-pixel locations from 42 raster scans. Afterward, we performed a double Gaussian fitting to the \ion{Si}{4} window. We limit our detection to those pixels where the FWHM values of both the regular and downflow components are greater than the thermal width of the spectral line (thermal FWHM, $v_{th} =\lambda/c\sqrt{8\ln(2)k_{B}T_{i}/m}$; where $\lambda$ -- wavelength of the line, $c$ -- speed of light, $k_{B}$ -- Boltzmann constant, $T_{i}$ --  ion-temperature, and $m$ -- ion-mass). If the peak values of both fitted Gaussian functions at one pixel have at least five counts, and both peaks are higher than 1$\sigma$ level, then we count it as a detected downflow. With this selection criterion, we have identified 4,976-pixel locations from the 42 observations. The locations of all the detected downflows from the sunspot observed on 2013 October 25 (the same sunspot shown in Figure~\ref{fig1}(A)) are marked in Figure~\ref{fig6}. The plot shows a big patch of downflows as well as a few isolated regions where downflows are present. The big patch might be part of a single extended downflow, whereas others might be independent of each other.  We examined our sample and found that around 14 rasters have extended (around 100 pixels or more) downflows. The distributions of the Doppler shifts of the regular component and the Doppler shifted component are also shown in the figure (using the detection from all the 42 raster observations). The scatter plot of the relationship between the Doppler velocity of the downflow component, and the solar-x is shown in the bottom right panel. It shows that these supersonic downflows generally do not have a strong dependence on the heliographic longitude on the solar disk.

After detection of downflows in the \ion{Si}{4} line, we focused on the full spectra as shown in the right panels of Figure~\ref{fig5}. The total spectra show six components, just like six well-isolated emission lines. Six Gaussian functions were simultaneously fitted to these three spectral lines and their satellite lines (due to strong downflows) at those selected pixel locations from the \ion{Si}{4} line. Similar to earlier detection criteria, we applied the following conditions: 1) the regular and secondary components should be separated by 40 km~s$^{-1}$ or larger. 2) After fitting, all the peak values of the six Gaussian functions must have at least 5 counts. 3) All the six peaks should be higher than 1$\sigma$ level. 4) The widths of all the six Gaussian functions should be higher than the thermal widths of the corresponding spectral lines. After applying this criterion, around 1,373-pixel locations from a total of 28 raster scans were selected. 
We should point out that the \ion{O}{4} lines generally have lower intensities than the \ion{Si}{4} line. For many of the locations where the detection is possible from the \ion{Si}{4} line, it is not possible to fit a 6-component Gaussian function properly to the spectra since the \ion{O}{4} signals are too low. This is the primary reason for having a lower pixel number in the final level of detection. Figure~\ref{fig5} shows the fitting results for the spectra of three different events. 
\begin{figure*}
\centering
\includegraphics[angle=00,clip,width=18.1cm]{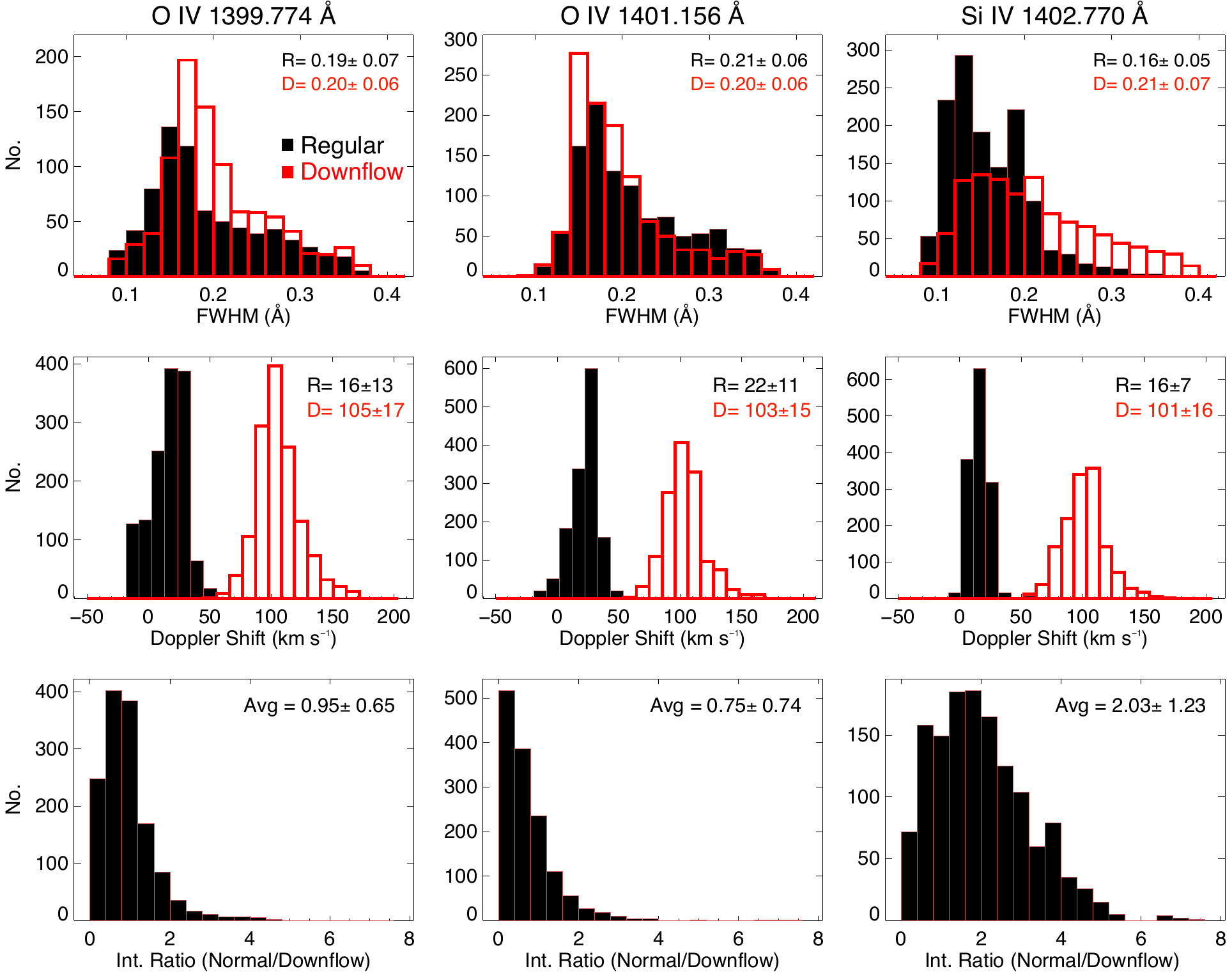}
\caption{Top panels: Distributions of the FWHM of the two doublets \ion{O}{4} and \ion{Si}{4} lines. Middle panels: Distributions of the Doppler velocity. Bottom panels: Distributions of the intensity ratio of the regular and downflow components. The black and red histograms represent the results for the regular and downflow components, respectively. The average value of each parameter with 1$\sigma$ error is also printed in each panel.}
\label{fig7} 
\end{figure*}
\begin{figure*}
\centering
\includegraphics[angle=00,clip,width=18.6cm]{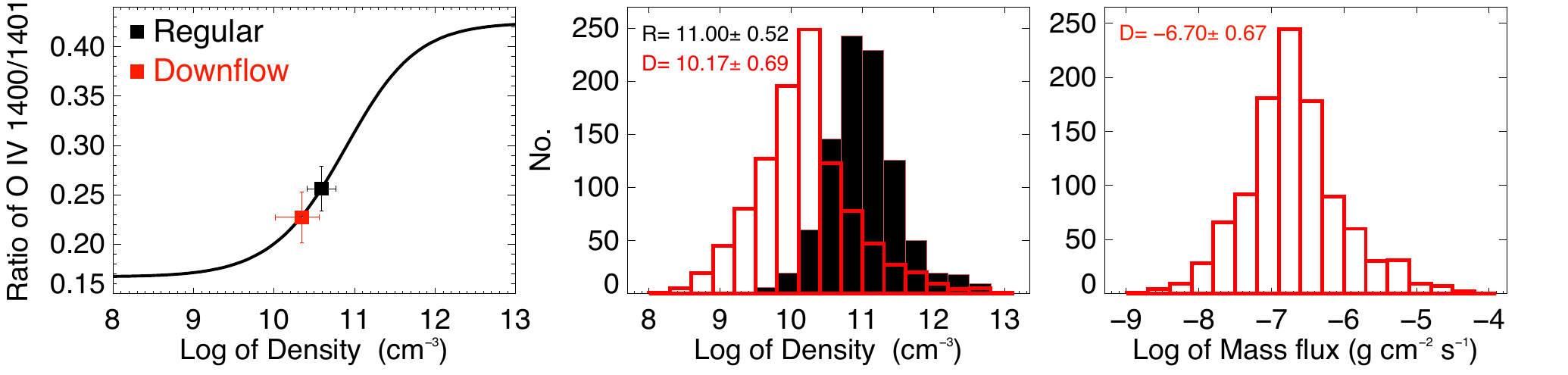}
\caption{ Left: Example of electron density diagnostics. The black curve shows the dependence of electron density on the intensity ratio of the \ion{O}{4} double lines. The black and red squares above the black curve mark the locations of the intensity ratio values that are obtained from Gaussian fitting of the spectral line profiles shown in Figure~\ref{fig5} (B).  Middle: The distribution of the measured density. The black and red histograms represent the density distributions of the regular and downflow components, respectively. The right panel shows the distribution of the mass flux due to supersonic downflows. The average value of each parameter with 1$\sigma$ error is printed in the middle and right panels.}
\label{fig8} 
\end{figure*}
\subsection{Statistical properties of the line parameters} 
Histograms of the FWHM and Doppler shift of both the regular component and the downflow component of the three lines are shown in the top and middle panels of Figure~\ref{fig7}. The distributions of the FWHM for the regular (R) and downflow (D) components of the \ion{O}{4} lines are very similar and have almost the same average values. Whereas the average FWHM of the \ion{Si}{4} line appears to be slightly larger in the downflow component than in the regular component. 
The distributions of the Doppler velocities derived from the six-component Gaussian fitting are presented in the middle panels, which show two distinct well-separated distributions. The regular component has a narrower distribution of velocity, whereas the supersonic downflow component has a broader velocity distribution. The average values of the supersonic downflow velocities of the \ion{O}{4} 1400/1401~{\AA} and \ion{Si}{4} 1403~{\AA} lines appear to be very similar ($\sim$ 100 km~s$^{-1}$). 
The intensity ratio between the regular and downflow components shows a broad distribution. The average value of this ratio is 0.95$\pm$0.65 for the \ion{O}{4} 1400~{\AA} line and 0.75$\pm$0.74 for the \ion{O}{4} 1401~{\AA} line, whereas the average ratio is 2.03$\pm$1.23 for the \ion{Si}{4} line.  

\subsection{Density and Mass flux} 
The \ion{O}{4} 1400/1401~{\AA} lines are very useful for density diagnostics of the TR, as the intensity ratio of the two lines is sensitive to electron density \citep[e.g.,][]{1992SoPh..138..283D,2009A&A...505..307T}. The \ion{O}{4} 1401~{\AA}  line is blended with the \ion{S}{1} 1401~{\AA} line, but the latter is generally very small within sunspots \citep{2002MNRAS.337..901K}. We used the CHIANTI atomic database version 7.1 \citep{1997A&AS..125..149D,2013ApJ...763...86L}  to compute the electron density from the \ion{O}{4} 1400~{\AA} and 1401~{\AA} line pair. The densities of both the regular and downflow components are determined using this intensity ratio method. We show the statistical distribution of the derived densities in Figure~\ref{fig8}.
The average value of the log of density for the regular component is 11.00$\pm$0.52, whereas the density of the supersonic downflow component has a much lower average value of 10.17$\pm$0.69. 

Using the derived density of the supersonic downflow at each pixel location, we estimated the mass flux associated with these downflows. The mass flux ($\rho v_{D}$) can be calculated as $n_{e} m_{p} (n_{H}/n_{e}) v_{D}$, where $\rho$ is the density,  $m_{p}$ is the mass of proton and  $n_{H}/n_{e}$  is the hydrogen to electron number density ratio (we assume this value to be 0.83). The velocity $v_{D}$  is taken from the measurement of the supersonic downflow velocity of the \ion{O}{4} 1401~{\AA} line, since this line generally has a higher intensity value than the \ion{O}{4} 1400~{\AA}  line and thus provides better fitting results. The histogram of the mass flux of the downflow is shown in Figure~\ref{fig8}. The average mass flux computed from the 28 observations is 10$^{-6.70\pm0.67}$ g~cm$^{-2}$~s$^{-1}$.

\subsection{Comparison between two types of downflows} 
\begin{figure*}
\centering
\includegraphics[angle=00,clip,width=18.4cm]{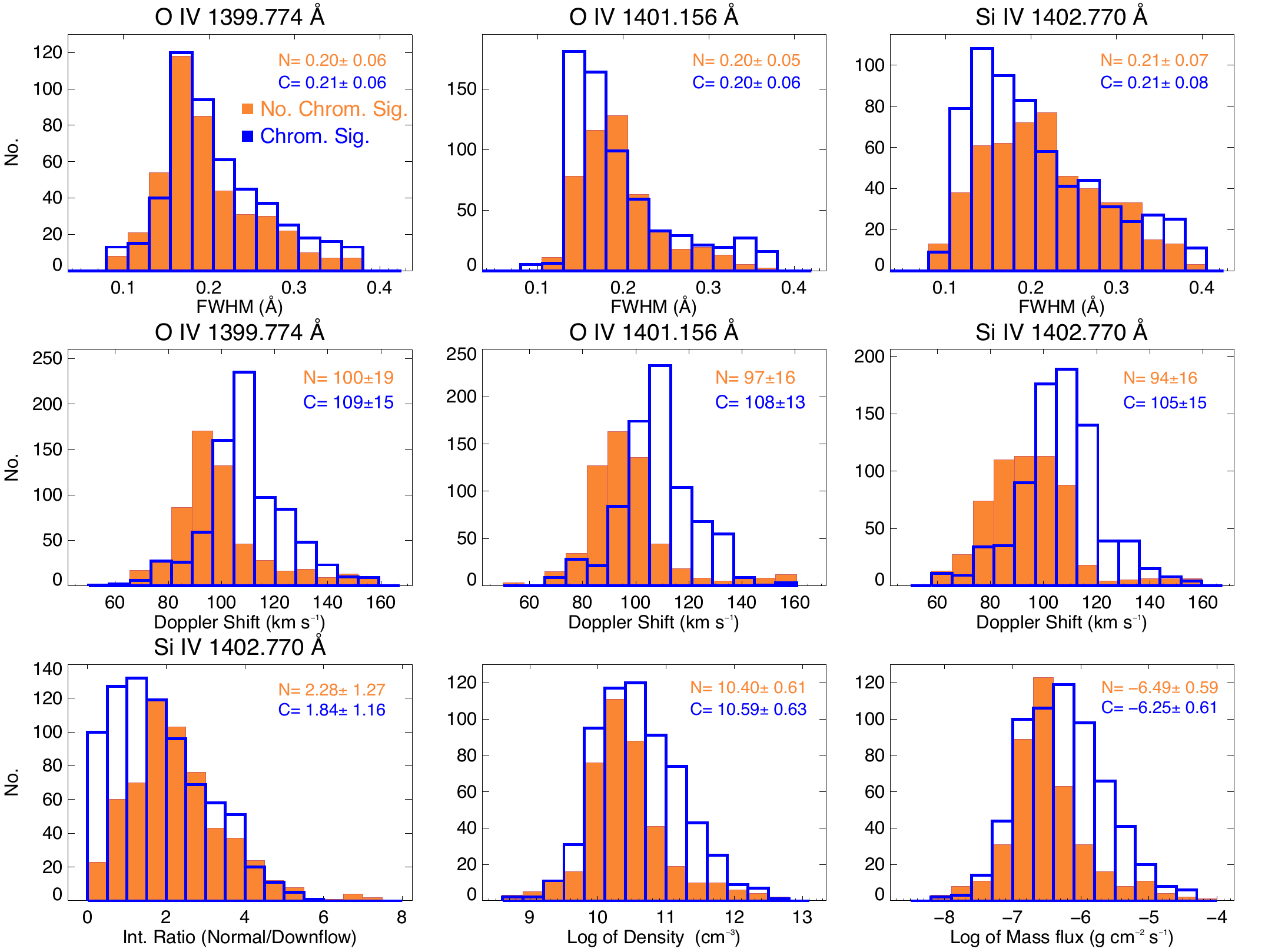}
\caption{Top panels: distributions of the FWHM of the two doublets \ion{O}{4} and \ion{Si}{4} lines for the supersonic downflows. Middle panels: distributions of the Doppler velocity. Bottom panels: the left panel shows the distribution of the intensity ratio of the regular and downflow components in the \ion{Si}{4} line, the middle panel shows the distribution of the density, and the right panel shows the distribution of the mass flux due to downflows. The orange histograms represent the results for the downflows without chromospheric signatures, whereas the blue represent the downflows with chromospheric counterparts. The average value of each parameter with 1$\sigma$ error is also printed in each panel.}
\label{fig9} 
\end{figure*}

As mentioned earlier, among 48 observations, 22 show signs of downflow in the chromospheric lines (a clear Doppler-shifted component in either the C II or Mg II lines or both), while others do not. 
After applying a 6-component Gaussian fitting to the spectra and determining the physical parameters of the downflows 
as described in the earlier sections, we have separated these two types of downflows. Histograms of the FWHM, Doppler shift, intensity ratio, density and mass flux for both types are presented in the Figure~\ref{fig9}. The distributions of the FWHM for the downflows with chromospheric counterparts and without chromospheric signatures are very similar and have almost similar average values. 
The distributions of the Doppler shift show that the downflows with chromospheric counterparts are slightly shifted towards higher values for the \ion{O}{4} doublet lines and \ion{Si}{4} line. This is also reflected in the average values of the distributions. The downflows with chromospheric counterparts show around 10 km s$^{-1}$ higher velocities in the average values of all the three lines.  
The intensity ratio between the regular and downflow components of the \ion{Si}{4} line shows a broad distribution for both types, though the distribution of downflows with chromospheric counterparts is shifted towards a smaller value. The average value of this ratio is 2.28$\pm$1.27 for downflows without chromospheric signatures and 1.84$\pm$1.16 for downflows with chromospheric counterparts. The distributions of the density and mass flux both have a small shift towards higher values for the downflows with chromospheric counterparts. The average values also show these differences. However, we must point out that these differences are within the 1$\sigma$ deviation of the average values of the distributions.

\section{Discussion}

In this work, we have performed a statistical study of sunspot supersonic downflows which are often seen in the TR lines \citep[e.g.,][]{2001ApJ...552L..77B, 2004ApJ...612.1193B,2014ApJ...786..137T,2015A&A...582A.116S,2016A&A...587A..20C}. As mentioned earlier,  \citet{2004ApJ...612.1193B} studied 12 sunspots using the SUMER data and found dual flows in only five of them. The fast downflow component and the regular component are usually overlapping with each other in the SUMER spectra. This is simply because the SUMER data has a comparatively low spatial resolution (about 2$''$--3$''$), much worse than that of IRIS (about 0.33$''$). Also, the instrument broadening of IRIS is much smaller than that of SUMER. As a result, the two components are often found to be well separated in the IRIS spectra, just like two well-separated emission lines. This allows us to accurately determine the line parameters of the two components by performing double Gaussian fitting. IRIS have observed many sunspots during the past few years. Using over 1.5-year sunspot raster data, we have performed a  comprehensive statistical study of these supersonic downflows and found that they are very common phenomena in sunspots. 

A total of 60 sunspot raster data were examined, and we found that among them 48 observations show supersonic downflows:  28 mostly occurring in the penumbrae, four in the umbrae and 16 show signatures of downflows both in the umbrae and penumbrae. With a much smaller sample, \cite{2004ApJ...612.1193B} and \cite{1993ApJ...412..865G} concluded earlier that the dual flows mostly occur in the penumbral regions. With the help of the high-resolution and high-sensitive IRIS instrument, we found that the supersonic downflows are very common not only in the penumbrae but also in the umbrae. Most of the sunspots which do not show a noticeable signature of such downflows are very small and isolated. 

\begin{figure*}[ht]
\centering
\includegraphics[angle=00,clip,width=18.4cm]{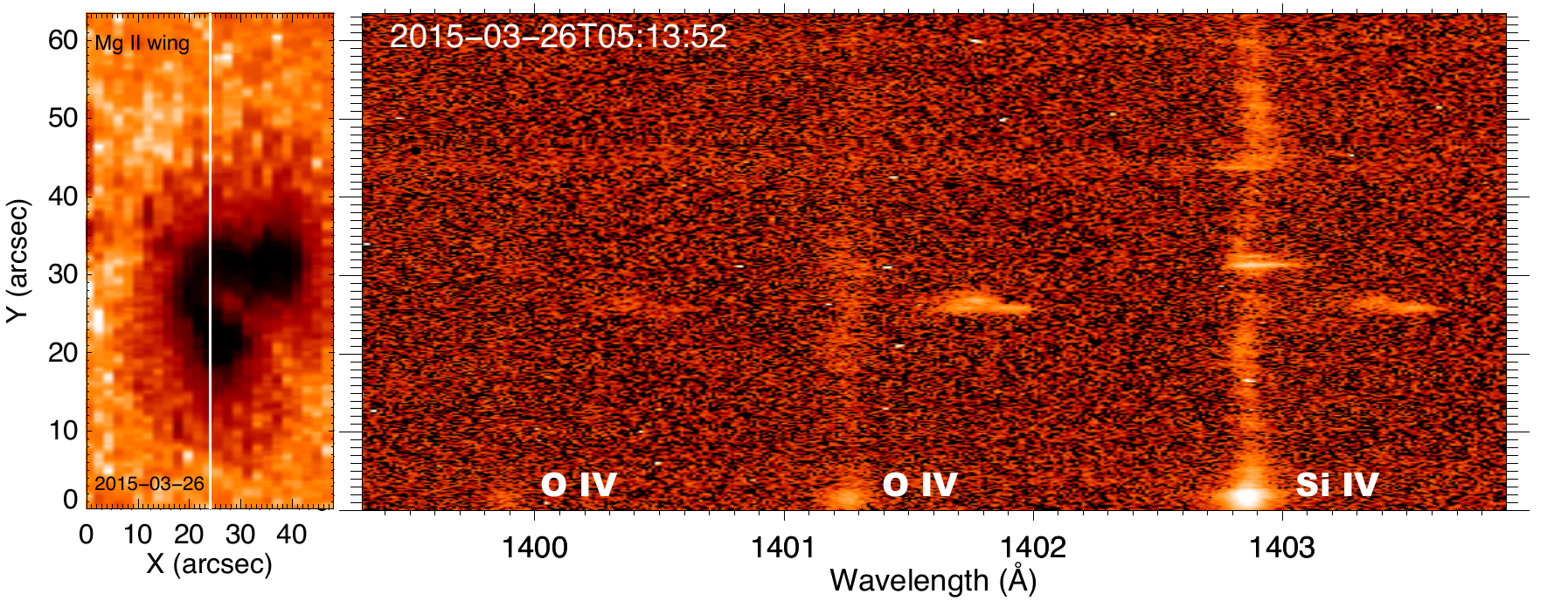}
\caption{Left panel shows \ion{Mg}{2}~k wing intensity image of a sunspot observed on  2015 March 16.  
A small window of the spectra along the solid line is shown in the right panel. The spectral window contains  \ion{O}{4} 1400/1401~{\AA} and \ion{Si}{4}1403~{\AA} lines. A supersonic downflow event is clearly seen in the spectra. It is also clear that at the location of the downflow, the regular component does not have significant intensity enhancement. }
\label{fig10} 
\end{figure*}

We have focused on the doublet  \ion{O}{4} 1400/1401~{\AA} and \ion{Si}{4} 1403~{\AA} lines for comprehensive spectroscopic analysis of the downflows. The statistical analysis over a large sample shows that the FWHM values of the regular and downflow components of the \ion{O}{4} lines are similar. Whereas on average the width of the \ion{Si}{4} line is larger in the downflow component than in the regular component. A similar result was also reported in a single event by \citet{2015A&A...582A.116S}. They found that the ratio of the FWHM of the downflow component and the corresponding main component for the \ion{Si}{4} line is 1.35,  and that this ratio is very close to 1 for the \ion{O}{4} lines. These ratios match well with our statistical average values. We find that the ratio of the FWHM between the downflow component and the corresponding regular component of the \ion{Si}{4} line has an average value of 1.3, whereas for the \ion{O}{4} 1400~{\AA} and  \ion{O}{4} 1401~{\AA} lines the average ratio is 1.05 and 0.95, respectively. 

The TR lines are mostly red-shifted, and there is no exception in sunspots \citep[e.g.,][]{2004A&A...428..629M,2008ApJ...685.1262M,2008AnGeo..26.2955D}. We found that the regular components of the typical TR lines are red-shifted on average. The average values of redshift are 16 km~s$^{-1}$ for the \ion{O}{4} 1400~{\AA} line, 22 km~s$^{-1}$ for the \ion{O}{4} 1400~{\AA} line and 16 km~s$^{-1}$ for the \ion{Si}{4} 1403~{\AA} line. The average redshifts of the secondary components are 105 km~s$^{-1}$, 103 km~s$^{-1}$ and 101 km~s$^{-1}$ for the same three lines, respectively. The average velocities of the \ion{O}{4}  and \ion{Si}{4} lines appear to be similar.  In the sit-and-stare observation of \citet{2015A&A...582A.116S} and the raster observation of \citet{2016A&A...587A..20C}, the authors found a $\sim$10 km~s$^{-1}$ difference in the velocities of the supersonic downflows seen in the \ion{Si}{4} 1403~{\AA} and \ion{O}{4} 1401~{\AA} lines. However, our statistical results do not show this difference. \citet{1990Ap&SS.170..135B} studied many TR region lines in the HRTS spectra and found that there is no strong dependence of the flow speed on temperature. Further studies are needed to have a clear and proper understanding of the temperature dependence of these supersonic downflows. 
 
The intensity ratios between the regular ($I_{R}$)  and downflow ($I_{D}$) components of the three lines show broad distributions. The average values for the \ion{O}{4} 1400~{\AA} and \ion{O}{4} 1401~{\AA} lines are around $I_{D}/I_{R}$ 0.95$\pm$0.65 and 0.75$\pm$0.74, respectively. While the ratio for the \ion{Si}{4} line is around 2.03$\pm$1.23. The large values of standard deviation might indicate that the intensities of the regular and downflow components are somewhat independent of each other. 
We also found that at some locations the downflow component is much stronger than the regular component. One example of such events is shown in the Figure~\ref{fig10}. 
The spectra show that the downflow component is very strong, and there is no significant enhancement of the regular component compared to the neighboring pixels at the location of the downflow. A similar effect can also be seen in Figure~\ref{fig5} C:  just above the dashed line a strong downflow component without significant  enhancement of the regular component is clearly seen. This might suggest that these high-speed downflowing materials are independent of the background TR plasma in sunspots.

We found that the electron density of the downflow component is generally lower than that of the regular component. The average value of the density for the regular component is $10^{11.00 \pm 0.52}$~cm$^{-3}$, whereas the density of the supersonic downflow component has a much lower average value of $10^{10.17 \pm 0.69}$~cm$^{-3}$.  
\citet{2015A&A...582A.116S} has also derived the electron density of a supersonic downflow event. They found that the density of the downflow component is $10^{10.6 \pm 0.25}$~cm$^{-3}$, which is only a factor 2 smaller than that of the regular component $10^{10.95 \pm 0.20}$~cm$^{-3}$.  However, our analysis based on a large sample shows that the average density value of the downflow component is around 7 times lower than that of the regular component.  This might indicate that these supersonic downflows are coming from lower-density regions, probably at higher layers in the corona.
Our analysis also shows that the mass flux due to supersonic downflows is very high. The mass flux obtained by \citet{2015A&A...582A.116S} is around 5 $\times$ 10$^{-7}$ g~cm$^{-2}$~s$^{-1}$, whereas we find a lower average mass flux of around 10$^{-6.70}$ g~cm$^{-2}$~s$^{-1}$ (or 2 $\times$ 10$^{-7}$ g~cm$^{-2}$~s$^{-1}$).
The IRIS sit-and-stare investigation of \citet{2014ApJ...786..137T} and \citet{2015A&A...582A.116S} shows that the downflows are remarkably steady over the total observing period of around 80 min.
Following the calculation of \citet{2016A&A...587A..20C}, we find that a typical coronal loop of 200-300 Mm length hosting downflows of 100 km~s$^{-1}$ will drain out all of its plasma around 100~s.  \citet{2015A&A...582A.116S} and \citet{2016A&A...587A..20C} proposed that these persistent supersonic downflows are due to termination shocks of siphon flows. But, till now we have not found any signatures of the upflow regions of siphon flows. It might be possible that the upflows have different temperatures and thus one cannot observe strong upflows in TR lines. Future high-resolution observations with simultaneous, multi-wavelength imaging and spectroscopy are needed to understand these phenomena completely.  
It would also be interesting to perform numerical simulations to understand where and how these downflows originate. To our knowledge, there are no existing numerical models which can reproduce these observed downflow features. Our statistical results will provide an important observational constraint to future numerical studies.

Earlier observations generally show no sign of supersonic downflows in the chromosphere. However, our statistical investigation shows that around 46\% of these downflows leave a trail in chromospheric lines. This suggests that TR supersonic downflows may be the reason behind some of the commonly observed supersonic downflows in the chromosphere \citep{2005ESASP.596E..49A,2010A&A...520A..77X,2007A&A...462.1147L}. We found that the average Doppler shift is about 10 km s$^{-1}$ higher for downflows with chromospheric counterparts compared to the ones without chromospheric signatures. 
The intensity ratio between the regular and downflow components of the \ion{Si}{4} line is on average slightly smaller for the downflows with chromospheric counterparts, which indicates that the intensity of the downflow component is generally higher for the downflows with chromospheric counterparts. The distributions of the density and mass flux are slightly shifted towards higher values for the downflows with chromospheric signatures. These results might indicate that downflows originating from a dense medium and having a higher velocity could reach the chromosphere. Finally, we noticed that the source of supersonic downflows at the photosphere \citep{2009ApJ...704L..29L,2011ApJ...727...49L,2013A&A...557A..24V} is still unknown, and whether they have any link to TR supersonic downflows is subject to further investigations. 

\section{Summary}
We have performed a comprehensive statistical study of the TR supersonic downflows and found that they are very common phenomena in sunspots. Using sunspot raster observations in the first 1.5 years after the launch of IRIS, we found that these supersonic downflows are present in $\sim$80\% of our sample. These downflows are observed primarily in the TR lines. However, some of them also appear in the \ion{C}{2} and \ion{Mg}{2} lines. At some locations, the downflow component is very strong, and there is no significant intensity enhancement in the regular component, suggesting that the downflowing materials are probably independent of the background plasma in sunspots. The downflows do not show any significant variations in heliographic longitude on the solar disk. Most downflows appear in the penumbrae, and some are detected in the umbrae. We have designed a six-component Gaussian fitting algorithm to quantify the intensity, velocity and width of the Si IV 1403~{\AA} and O IV 1400/1401~{\AA} line profiles in these downflows. Using the intensity ratio method, we computed the electron density for both the regular and downflow components. We also found that on average the downflows with chromospheric counterparts have a $\sim$10 km s$^{-1}$ higher velocity and slightly higher density compared to those without chromospheric signatures.
A large sample of data suggests that the downflow component has a much lower density compared to the regular component. We have also computed the mass flux due to these supersonic downflows, which is 10$^{-6.70\pm0.67}$ g~cm$^{-2}$~s$^{-1}$ on average. Our statistical results will provide a constraint to future numerical studies of sunspot supersonic downflows.

\acknowledgments
We thank Hardi Peter and Bernhard Fleck for helpful discussion and suggestions, and Sean McKillop for developing an initial code of downflow detection. 
This work is supported by NSFC grants 41574166 and 11790304 (11790300), the Strategic Pioneer Program on Space Science, Chinese Academy of Sciences, grants XDA15011000 and XDA-17040505, the Recruitment Program of Global Experts of China, and the Max Planck Partner group program. DPC is supported by NSF (USA) grant 1413686. IRIS is a NASA Small Explorer mission developed and operated by LMSAL with mission operations executed at NASA Ames Research center and major contributions to downlink communications funded by ESA and the Norwegian Space Center. CHIANTI is a collaborative project involving George Mason University, the University of Michigan (USA) and the University of Cambridge (UK).
\bibliographystyle{apj}
\bibliography{references}

\begin{thebibliography}{}
\expandafter\ifx\csname natexlab\endcsname\relax\def\natexlab#1{#1}\fi

\bibitem[{{Alpert} {et~al.}(2016){Alpert}, {Tiwari}, {Moore}, {Winebarger}, \&
  {Savage}}]{2016ApJ...822...35A}
{Alpert}, S.~E., {Tiwari}, S.~K., {Moore}, R.~L., {Winebarger}, A.~R., \&
  {Savage}, S.~L. 2016, \apj, 822, 35

\bibitem[{{Aznar Cuadrado} {et~al.}(2005){Aznar Cuadrado}, {Solanki}, \&
  {Lagg}}]{2005ESASP.596E..49A}
{Aznar Cuadrado}, R., {Solanki}, S.~K., \& {Lagg}, A. 2005, in ESA Special
  Publication, Vol. 596, Chromospheric and Coronal Magnetic Fields, ed. D.~E.
  {Innes}, A.~{Lagg}, \& S.~A. {Solanki}, 49.1

\bibitem[{{Bai} {et~al.}(2016){Bai}, {Su}, {Cao}, {Liu}, {Deng}, \&
  {Priya}}]{2016ApJ...823...60B}
{Bai}, X.~Y., {Su}, J.~T., {Cao}, W.~D., {et~al.} 2016, \apj, 823, 60

\bibitem[{{Borrero} \& {Ichimoto}(2011)}]{2011LRSP....8....4B}
{Borrero}, J.~M., \& {Ichimoto}, K. 2011, Living Reviews in Solar Physics, 8, 4

\bibitem[{{Brekke} {et~al.}(1990){Brekke}, {Kjeldseth-Moe}, \&
  {Brueckner}}]{1990Ap&SS.170..135B}
{Brekke}, P., {Kjeldseth-Moe}, O., \& {Brueckner}, G.~E. 1990, \apss, 170, 135

\bibitem[{{Brynildsen} {et~al.}(2001){Brynildsen}, {Maltby}, {Kjeldseth-Moe},
  \& {Wilhelm}}]{2001ApJ...552L..77B}
{Brynildsen}, N., {Maltby}, P., {Kjeldseth-Moe}, O., \& {Wilhelm}, K. 2001,
  \apjl, 552, L77

\bibitem[{{Brynildsen} {et~al.}(2004){Brynildsen}, {Maltby}, {Kjeldseth-Moe},
  \& {Wilhelm}}]{2004ApJ...612.1193B}
---. 2004, \apj, 612, 1193

\bibitem[{{Chitta} {et~al.}(2016){Chitta}, {Peter}, \&
  {Young}}]{2016A&A...587A..20C}
{Chitta}, L.~P., {Peter}, H., \& {Young}, P.~R. 2016, \aap, 587, A20

\bibitem[{{Dammasch} {et~al.}(2008){Dammasch}, {Curdt}, {Dwivedi}, \&
  {Parenti}}]{2008AnGeo..26.2955D}
{Dammasch}, I.~E., {Curdt}, W., {Dwivedi}, B.~N., \& {Parenti}, S. 2008,
  Annales Geophysicae, 26, 2955

\bibitem[{{De Pontieu} {et~al.}(2014){De Pontieu}, {Title}, {Lemen}, {Kushner},
  {Akin}, {Allard}, {Berger}, {Boerner}, {Cheung}, {Chou}, {Drake}, {Duncan},
  {Freeland}, {Heyman}, {Hoffman}, {Hurlburt}, {Lindgren}, {Mathur}, {Rehse},
  {Sabolish}, {Seguin}, {Schrijver}, {Tarbell}, {W{\"u}lser}, {Wolfson},
  {Yanari}, {Mudge}, {Nguyen-Phuc}, {Timmons}, {van Bezooijen}, {Weingrod},
  {Brookner}, {Butcher}, {Dougherty}, {Eder}, {Knagenhjelm}, {Larsen},
  {Mansir}, {Phan}, {Boyle}, {Cheimets}, {DeLuca}, {Golub}, {Gates}, {Hertz},
  {McKillop}, {Park}, {Perry}, {Podgorski}, {Reeves}, {Saar}, {Testa}, {Tian},
  {Weber}, {Dunn}, {Eccles}, {Jaeggli}, {Kankelborg}, {Mashburn}, {Pust},
  {Springer}, {Carvalho}, {Kleint}, {Marmie}, {Mazmanian}, {Pereira}, {Sawyer},
  {Strong}, {Worden}, {Carlsson}, {Hansteen}, {Leenaarts}, {Wiesmann},
  {Aloise}, {Chu}, {Bush}, {Scherrer}, {Brekke}, {Martinez-Sykora}, {Lites},
  {McIntosh}, {Uitenbroek}, {Okamoto}, {Gummin}, {Auker}, {Jerram}, {Pool}, \&
  {Waltham}}]{2014SoPh..289.2733D}
{De Pontieu}, B., {Title}, A.~M., {Lemen}, J.~R., {et~al.} 2014, \solphys, 289,
  2733

\bibitem[{{Deng} {et~al.}(2016){Deng}, {Yurchyshyn}, {Tian}, {Kleint}, {Liu},
  {Xu}, \& {Wang}}]{2016ApJ...829..103D}
{Deng}, N., {Yurchyshyn}, V., {Tian}, H., {et~al.} 2016, \apj, 829, 103

\bibitem[{{Dere}(1982)}]{1982SoPh...77...77D}
{Dere}, K.~P. 1982, \solphys, 77, 77

\bibitem[{{Dwivedi} \& {Gupta}(1992)}]{1992SoPh..138..283D}
{Dwivedi}, B.~N., \& {Gupta}, A.~K. 1992, \solphys, 138, 283

\bibitem[{{Gurman}(1993)}]{1993ApJ...412..865G}
{Gurman}, J.~B. 1993, \apj, 412, 865

\bibitem[{{Harrison} {et~al.}(1995){Harrison}, {Sawyer}, {Carter}, \& {36
  co-authors}}]{Harrison95}
{Harrison}, R.~A., {Sawyer}, E.~C., {Carter}, M.~K., \& {36 co-authors}. 1995,
  \solphys, 162, 233


\bibitem[{{Keenan} {et~al.}(2002){Keenan}, {Ahmed}, {Brage}, {Doyle}, {Espey},
  {Exter}, {Hibbert}, {Keenan}, {Madjarska}, {Mathioudakis}, \&
  {Pollacco}}]{2002MNRAS.337..901K}
{Keenan}, F.~P., {Ahmed}, S., {Brage}, T., {et~al.} 2002, \mnras, 337, 901

\bibitem[{{Kjeldseth-Moe} {et~al.}(1988){Kjeldseth-Moe}, {Brynildsen},
  {Brekke}, {Engvold}, {Maltby}, {Bartoe}, {Brueckner}, {Cook}, {Dere}, \&
  {Socker}}]{1988ApJ...334.1066K}
{Kjeldseth-Moe}, O., {Brynildsen}, N., {Brekke}, P., {et~al.} 1988, \apj, 334,
  1066

\bibitem[{{Kleint} {et~al.}(2014){Kleint}, {Antolin}, {Tian}, {Judge}, {Testa},
  {De Pontieu}, {Mart{\'{\i}}nez-Sykora}, {Reeves}, {Wuelser}, {McKillop},
  {Saar}, {Carlsson}, {Boerner}, {Hurlburt}, {Lemen}, {Tarbell}, {Title},
  {Golub}, {Hansteen}, {Jaeggli}, \& {Kankelborg}}]{2014ApJ...789L..42K}
{Kleint}, L., {Antolin}, P., {Tian}, H., {et~al.} 2014, \apjl, 789, L42

\bibitem[{{K.P. Dere} {et~al.}(1997){K.P. Dere}, {E. Landi}, {H.E. Mason},
  {B.C. Monsignori Fossi}, \& {P.R. Young}}]{1997A&AS..125..149D}
{K.P. Dere}, {E. Landi}, {H.E. Mason}, {B.C. Monsignori Fossi}, \& {P.R.
  Young}. 1997, Astron. Astrophys. Suppl. Ser., 125, 149

\bibitem[Lagg et al.(2007)]{2007A&A...462.1147L} 
Lagg, A., Woch, J., Solanki, S.~K., \& Krupp, N.\ 2007, \aap, 462, 1147 

\bibitem[{{Landi} {et~al.}(2013){Landi}, {Young}, {Dere}, {Del Zanna}, \&
  {Mason}}]{2013ApJ...763...86L}
{Landi}, E., {Young}, P.~R., {Dere}, K.~P., {Del Zanna}, G., \& {Mason}, H.~E.
  2013, \apj, 763, 86

\bibitem[{{Lemen} {et~al.}(2012){Lemen}, {Title}, {Akin}, {Boerner}, {Chou},
  {Drake}, {Duncan}, {Edwards}, {Friedlaender}, {Heyman}, {Hurlburt}, {Katz},
  {Kushner}, {Levay}, {Lindgren}, {Mathur}, {McFeaters}, {Mitchell}, {Rehse},
  {Schrijver}, {Springer}, {Stern}, {Tarbell}, {Wuelser}, {Wolfson}, {Yanari},
  {Bookbinder}, {Cheimets}, {Caldwell}, {Deluca}, {Gates}, {Golub}, {Park},
  {Podgorski}, {Bush}, {Scherrer}, {Gummin}, {Smith}, {Auker}, {Jerram},
  {Pool}, {Soufli}, {Windt}, {Beardsley}, {Clapp}, {Lang}, \&
  {Waltham}}]{2012SoPh..275...17L}
{Lemen}, J.~R., {Title}, A.~M., {Akin}, D.~J., {et~al.} 2012, \solphys, 275, 17

\bibitem[{{Louis} {et~al.}(2009){Louis}, {Bellot Rubio}, {Mathew}, \&
  {Venkatakrishnan}}]{2009ApJ...704L..29L}
{Louis}, R.~E., {Bellot Rubio}, L.~R., {Mathew}, S.~K., \& {Venkatakrishnan},
  P. 2009, \apjl, 704, L29

\bibitem[{{Louis} {et~al.}(2011){Louis}, {Bellot Rubio}, {Mathew}, \&
  {Venkatakrishnan}}]{2011ApJ...727...49L}
---. 2011, \apj, 727, 49

\bibitem[{{Marsch} {et~al.}(2004){Marsch}, {Wiegelmann}, \&
  {Xia}}]{2004A&A...428..629M}
{Marsch}, E., {Wiegelmann}, T., \& {Xia}, L.~D. 2004, \aap, 428, 629

\bibitem[Marsch et al.(2008)]{2008ApJ...685.1262M} 
Marsch, E., Tian, H., Sun, J., Curdt, W., \& Wiegelmann, T.\ 2008, \apj, 685, 1262-1269

\bibitem[{{Nicolas} {et~al.}(1982){Nicolas}, {Bartoe}, {Brueckner}, \&
  {Kjeldseth-Moe}}]{1982SoPh...81..253N}
{Nicolas}, K.~R., {Bartoe}, J.-D.~F., {Brueckner}, G.~E., \& {Kjeldseth-Moe},
  O. 1982, \solphys, 81, 253

\bibitem[{{Reid} {et~al.}(2018){Reid}, {Henriques}, {Mathioudakis}, \&
  {Samanta}}]{2018ApJ...855L..19R}
{Reid}, A., {Henriques}, V.~M.~J., {Mathioudakis}, M., \& {Samanta}, T. 2018,
  \apjl, 855, L19

\bibitem[{{Samanta} {et~al.}(2017){Samanta}, {Tian}, {Banerjee}, \&
  {Schanche}}]{2017ApJ...835L..19S}
{Samanta}, T., {Tian}, H., {Banerjee}, D., \& {Schanche}, N. 2017, \apjl, 835,
  L19

\bibitem[{{Solanki}(2003)}]{2003A&ARv..11..153S}
{Solanki}, S.~K. 2003, \aapr, 11, 153


\bibitem[{{Straus} {et~al.}(2015){Straus}, {Fleck}, \&
  {Andretta}}]{2015A&A...582A.116S}
{Straus}, T., {Fleck}, B., \& {Andretta}, V. 2015, \aap, 582, A116

\bibitem[{{Tian} {et~al.}(2014{\natexlab{b}}){Tian}, {DeLuca}, {Reeves},
  {McKillop}, {De Pontieu}, {Mart{\'{\i}}nez-Sykora}, {Carlsson}, {Hansteen},
  {Kleint}, {Cheung}, {Golub}, {Saar}, {Testa}, {Weber}, {Lemen}, {Title},
  {Boerner}, {Hurlburt}, {Tarbell}, {Wuelser}, {Kankelborg}, {Jaeggli}, \&
  {McIntosh}}]{2014ApJ...786..137T}
{Tian}, H., {DeLuca}, E., {Reeves}, K.~K., {et~al.} 2014{\natexlab{b}}, \apj,
  786, 137

\bibitem[{{Tian} {et~al.}(2009){Tian}, {Curdt}, {Teriaca}, {Landi}, \&
  {Marsch}}]{2009A&A...505..307T}
{Tian}, H., {Curdt}, W., {Teriaca}, L., {Landi}, E., \& {Marsch}, E. 2009,
  \aap, 505, 307

\bibitem[{{Tian} {et~al.}(2014{\natexlab{a}}){Tian}, {Kleint}, {Peter},
  {Weber}, {Testa}, {DeLuca}, {Golub}, \& {Schanche}}]{2014ApJ...790L..29T}
{Tian}, H., {Kleint}, L., {Peter}, H., {et~al.} 2014{\natexlab{a}}, \apjl, 790,
  L29

\bibitem[{{Tian}(2017)}]{2017RAA....17..110T}
{Tian}, H. 2017, Research in Astronomy and Astrophysics, 17, 110

\bibitem[{Tian {et~al.}(2018b)Tian, Samanta, \& Zhang}]{Tian2018}
Tian, H., Samanta, T., \& Zhang, J. 2018b, Geoscience Letters, 5, 4

\bibitem[{{Tian} {et~al.}(2018a){Tian}, {Yurchyshyn}, {Peter}, {Solanki},
  {Young}, {Ni}, {Cao}, {Ji}, {Zhu}, {Zhang}, {Samanta}, {Song}, {He}, {Wang},
  \& {Chen}}]{2018ApJ...854...92T}
{Tian}, H., {Yurchyshyn}, V., {Peter}, H., {et~al.} 2018a, \apj, 854, 92

\bibitem[Toriumi et al.(2015)]{2015ApJ...811..137T} 
Toriumi, S., Katsukawa, Y., \& Cheung, M.~C.~M.\ 2015, \apj, 811, 137

\bibitem[{{Tiwari}(2017)}]{2017arXiv171207174T}
{Tiwari}, S.~K. 2017, ArXiv e-prints, arXiv:1712.07174

\bibitem[{{van Noort} {et~al.}(2013){van Noort}, {Lagg}, {Tiwari}, \&
  {Solanki}}]{2013A&A...557A..24V}
{van Noort}, M., {Lagg}, A., {Tiwari}, S.~K., \& {Solanki}, S.~K. 2013, \aap,
  557, A24

\bibitem[{{Vissers} {et~al.}(2015){Vissers}, {Rouppe van der Voort}, \&
  {Carlsson}}]{2015ApJ...811L..33V}
{Vissers}, G.~J.~M., {Rouppe van der Voort}, L.~H.~M., \& {Carlsson}, M. 2015,
  \apjl, 811, L33

\bibitem[{{Wilhelm} {et~al.}(1995){Wilhelm}, {Curdt}, {Marsch}, {Sch{\"u}hle},
  {Lemaire}, {Gabriel}, {Vial}, {Grewing}, {Huber}, {Jordan}, {Poland},
  {Thomas}, {K{\"u}hne}, {Timothy}, {Hassler}, \&
  {Siegmund}}]{1995SoPh..162..189W}
{Wilhelm}, K., {Curdt}, W., {Marsch}, E., {et~al.} 1995, \solphys, 162, 189

\bibitem[{{Xu} {et~al.}(2010){Xu}, {Lagg}, \& {Solanki}}]{2010A&A...520A..77X}
{Xu}, Z., {Lagg}, A., \& {Solanki}, S.~K. 2010, \aap, 520, A77


\end{thebibliography}

\appendix


\begin{table}[ht]
\begin{center}
\caption{List of data and the details of SuperSonic Downflows (SSDs)}
\begin{tabular}{cccccccc}
\hline
\hline

\multicolumn{1}{c}{Active Region }&\multicolumn{1}{c}{Date \& Starting Time}&\multicolumn{1}{c}{Visual Detection}&\multicolumn{1}{c}{Chromospheric } &\multicolumn{1}{c}{SSDs location}&\multicolumn{1}{c}{SSDs location}&\multicolumn{1}{c}{Automated Detection}&\multicolumn{1}{c}{6-Gaussian fitting }\\
\multicolumn{1}{c}{Number}&\multicolumn{1}{c}{of the Raster}&\multicolumn{1}{c}{of SSDs in \ion{Si}{4}} &\multicolumn{1}{c}{Signature}&\multicolumn{1}{c}{(Umbra)}&\multicolumn{1}{c}{(Penumbra)}&\multicolumn{1}{c}{of SSDs in \ion{Si}{4}}&\multicolumn{1}{c}{applied}\\

\hline
\hline
11836        & 2013-09-02 06:29:35          &\xmark     &\xmark     &\xmark       &\xmark     &\xmark       & \xmark \\
11841        & 2013-09-17 19:30:43          &\xmark     &\xmark     &\xmark       &\xmark     &\xmark       & \xmark \\
11843            & 2013-09-19 19:53:03          &\xmark     &\xmark     &\xmark       &\xmark     &\xmark       & \xmark \\
11846            & 2013-09-21 14:31:37          &\cmark     &\cmark     &\cmark       &\cmark     &\cmark       & \cmark \\
11861            & 2013-10-12 20:01:44          &\cmark     &\xmark     &\xmark       &\cmark     &\cmark       & \cmark \\
11877            & 2013-10-25 05:05:30          &\cmark     &\cmark     &\cmark       &\cmark     &\cmark       & \cmark \\
11899            & 2013-11-20 14:11:51          &\cmark     &\xmark     &\xmark       &\cmark     &\cmark       & \xmark \\
11903            & 2013-11-23 15:36:09          &\cmark     &\cmark     &\xmark       &\cmark     &\cmark       & \cmark \\
11907            & 2013-11-29 09:46:04          &\cmark     &\xmark     &\xmark       &\cmark     &\cmark       & \cmark \\
11908            & 2013-11-30 18:20:24          &\cmark     &\cmark     &\xmark       &\cmark     &\cmark       & \xmark \\
11908            & 2013-11-30 20:00:24          &\cmark     &\cmark     &\xmark       &\cmark     &\cmark       & \cmark \\
11909            & 2013-12-03 11:14:38          &\cmark     &\xmark     &\cmark       &\cmark     &\cmark       & \xmark \\
11916            & 2013-12-06 07:34:51          &\xmark     &\xmark     &\xmark       &\xmark     &\xmark       & \xmark \\
11930            & 2013-12-20 16:27:38          &\cmark     &\xmark     &\xmark       &\cmark     &\cmark       & \xmark \\
11931            & 2013-12-25 18:34:51          &\cmark     &\xmark     &\xmark       &\cmark     &\cmark       & \xmark \\
11941            & 2014-01-01 13:04:31          &\cmark     &\cmark     &\cmark       &\cmark     &\xmark       & \xmark \\
11944            & 2014-01-07 19:01:37          &\cmark     &\xmark     &\xmark       &\cmark     &\cmark       & \xmark \\
11944            & 2014-01-08 01:15:21          &\cmark     &\cmark     &\cmark       &\cmark     &\cmark       & \xmark \\
11959            & 2014-01-24 17:32:51          &\xmark     &\xmark     &\xmark       &\xmark     &\xmark       & \xmark \\
11967            & 2014-02-02 21:08:30          &\cmark     &\cmark     &\cmark       &\cmark     &\cmark       & \cmark \\
11974            & 2014-02-12 20:16:09          &\cmark     &\cmark     &\xmark       &\cmark     &\cmark       & \cmark \\
11982            & 2014-02-25 10:49:43          &\cmark     &\cmark     &\xmark       &\cmark     &\cmark       & \cmark \\
11990            & 2014-03-02 23:03:41          &\cmark     &\xmark     &\xmark       &\cmark     &\cmark       & \cmark \\
12014            & 2014-03-25 09:40:20          &\cmark     &\cmark     &\cmark       &\cmark     &\cmark       & \xmark \\
12045            & 2014-04-27 15:36:05          &\cmark     &\xmark     &\xmark       &\cmark     &\xmark       & \xmark \\
12049            & 2014-05-03 15:52:14          &\cmark     &\xmark     &\xmark       &\cmark     &\cmark       & \cmark \\
12056            & 2014-05-11 18:32:01          &\cmark     &\xmark     &\xmark       &\cmark     &\xmark       & \xmark \\
12071            & 2014-05-23 05:04:58          &\cmark     &\xmark     &\cmark       &\cmark     &\cmark       & \cmark \\
12077            & 2014-06-04 08:03:09          &\cmark     &\xmark     &\cmark       &\cmark     &\cmark       & \cmark \\
12079            & 2014-06-08 05:13:58          &\cmark     &\cmark     &\cmark       &\cmark     &\cmark       & \cmark \\
12090            & 2014-06-18 18:10:04          &\cmark     &\xmark     &\cmark       &\cmark     &\cmark       & \xmark \\
12096            & 2014-06-28 10:08:48          &\cmark     &\cmark     &\xmark       &\cmark     &\cmark       & \xmark \\
12104            & 2014-07-04 11:40:30          &\cmark     &\cmark     &\xmark       &\cmark     &\xmark       & \xmark \\
12109            & 2014-07-09 04:10:00          &\cmark     &\cmark     &\cmark       &\xmark     &\cmark       & \cmark \\
12108            & 2014-07-09 07:21:05          &\cmark     &\xmark     &\xmark       &\cmark     &\xmark       & \xmark \\
12121            & 2014-07-29 05:15:00          &\cmark     &\xmark     &\cmark       &\xmark     &\cmark       & \cmark \\
12130            & 2014-08-02 14:06:41          &\cmark     &\xmark     &\cmark       &\cmark     &\cmark       & \cmark \\
12127            & 2014-08-03 13:30:01          &\cmark     &\xmark     &\cmark       &\cmark     &\xmark       & \xmark \\
12132            & 2014-08-05 13:30:03          &\cmark     &\cmark     &\xmark      &\cmark     &\cmark       & \cmark \\
12135            & 2014-08-13 05:01:52          &\cmark     &\cmark     &\cmark       &\xmark     &\cmark       & \cmark \\
12146            & 2014-08-21 21:41:51          &\cmark     &\xmark     &\xmark       &\cmark     &\cmark       & \cmark \\
12149        & 2014-08-28 21:06:01          &\xmark     &\xmark     &\xmark       &\xmark     &\xmark       & \xmark \\
12157        & 2014-09-10 06:57:51          &\xmark     &\xmark     &\xmark       &\xmark     &\xmark       & \xmark \\
12178            & 2014-10-03 14:42:58          &\cmark     &\xmark     &\cmark       &\xmark     &\cmark       & \cmark \\
12186            & 2014-10-14 17:07:00          &\cmark     &\xmark     &\xmark       &\cmark     &\cmark       & \cmark \\
12209        & 2014-11-19 15:47:08          &\xmark     &\xmark     &\xmark       &\xmark     &\xmark       & \xmark \\
12209        & 2014-11-19 17:24:20          &\xmark     &\xmark     &\xmark       &\xmark     &\xmark       & \xmark \\
12219            & 2014-11-28 19:34:46          &\cmark     &\cmark     &\xmark       &\cmark    &\cmark       & \cmark \\
12222            & 2014-12-02 05:00:34          &\cmark     &\cmark     &\cmark       &\cmark     &\cmark       & \xmark \\
12230            & 2014-12-11 17:59:04          &\cmark     &\xmark     &\xmark       &\cmark     &\cmark       & \cmark \\
12235        & 2014-12-15 17:29:21          &\xmark     &\xmark     &\xmark       &\xmark     &\xmark       & \xmark \\
12252        & 2015-01-01 19:40:18          &\cmark     &\xmark     &\xmark       &\cmark     &\cmark       & \cmark \\
12253        & 2015-01-07 13:25:15          &\xmark     &\xmark     &\xmark       &\xmark     &\xmark       & \xmark \\
12257        & 2015-01-10 01:33:15          &\cmark     &\xmark     &\xmark       &\cmark     &\cmark       & \xmark \\
12259        & 2015-01-13 10:51:18          &\xmark     &\xmark     &\xmark       &\xmark     &\xmark       & \xmark \\
12277        & 2015-02-05 06:49:19          &\cmark     &\xmark     &\xmark       &\cmark     &\cmark       & \cmark \\
12280            & 2015-02-06 05:15:07          &\cmark     &\cmark     &\xmark       &\cmark     &\cmark       & \xmark \\
12281            & 2015-02-08 05:59:16          &\cmark     &\cmark     &\xmark       &\cmark     &\cmark       & \xmark \\
12282            & 2015-02-14 15:04:07         &\cmark     &\xmark     &\cmark       &\cmark     &\cmark       & \cmark \\
12305            & 2015-03-26 04:59:26             &\cmark     &\cmark     &\cmark       &\cmark     &\cmark       & \cmark \\

\hline
\hline
\end{tabular}
\label{tab10}
\end{center}
\end{table}

\end{document}